\documentclass[epj]{svjour}
\usepackage{graphics}
\usepackage{amsmath}
\usepackage{latexsym}
\begin{document}
\title{Fermions tunnelling with quantum gravity correction}

\author{Zhen-Yu liu\inst{1}£¬ \and Ji-Rong Ren\inst{2}
\thanks{\emph{email:} renjr@lzu.edu.cn, corresponding author. }%
}                     
%
%
\institute{Institute of Theoretical Physics, Lan zhou University, lanzhou, 73000, China}
\date{Received: date / Revised version: date}
%
\abstract{Quantum gravity correction is truly important to
 study  tunnelling process of black hole.
 Base on the generalized uncertainty principle,
 we investigate the influence of quantum gravity and
 the result tell us that the quantum gravity correction accelerates the evaporation of black hole.
 Using corrected Dirac equation in curved spacetime and Hamilton-Jacobi method, we address the tunnelling of fermions in a 4-dimensional Schwarzschild spacetime.
 After solving the equation of motion of the spin 1/2 field, we obtain the corrected Hawking temperature.
 It turns out that the correction depends not only on the  mass of black hole but aslo on the mass of emitted fermions.
 In our calculation, the quantum gravity correction accelerates the increasing of Hawking temperature during the radiation explicitly.
 This correction leads to the increasing of the evaporation of black hole.
\PACS{
     {0470Dy}{Quantum aspects of black holes, evaporation, thermodynamics}   \and
     {0460-m}{Quantum gravity}
     } 
} 
\maketitle
\section{Introduction}
\label{intro}
Hawking radiation is a process of quantum tunnelling of particles at black hole's horizons\cite{HaR}. Kraus and Wilczek first discussed the radiation in the view of the dynamics of space-time\cite{PKFW}. They studied the massless scalar particle's Hawking radiation as a tunnelling process based on a dynamical geometry\cite{MPFW}. In their work, the modified emission spectra for spherically symmetric black holes and the corrected Hawking temperature were derived. And the tunnelling rate is connected with the change of Bekenstein-Hawking entropy. As for general black holes, Hawking radiation was studied in\cite{JZZ,HRKN}. The fermions tunnelling from black holes was studied with Hamilton-Jacobi method and the Hawking temperature was recoverd\cite{FTBH}. Fermions tunnelling from the outer horizon of BTZ black hole is investigated in\cite{FTBTZ}. More about fermions tunnelling investigations in complicated spacetimes are referred to\cite{FTRDS,FTGHSD,FTDYH,FTHB}.

Most approaches to quantum gravity theories predict the existence of the minimal length\cite{PKGC,DMG,KGP,LGML,RPL}. In the theoretical framework, the minimal length can be achieved in different ways\cite{SHN}.  One way to realize the minimal length is utilizing the generalized uncertainty principle (GUP). Kempf et al.\cite{KEM,AKE} proposed a
three-dimensional generalized uncertainty relation of the form
\begin{equation}
[\hat{X_i},\hat{P_j}]=i\hbar\left[(1+\beta \hat{\bf P}^2)\delta_{ij}+\beta^{'}\hat{P_i}\hat{P_j}\right]
\end{equation}
where $\beta$ and $\beta^{'}$ are parameters related to the minimal length. If assume $\beta^{'}=2\beta$ and $[\hat{P_i},\hat{P_j}]=0$, then to the first order of $\beta$, the form of the modified position operator and momentum operator will be
\begin{equation}
\hat{X_i}=x_{i}, \quad
\hat{P}_i=p_{i}(1+\beta p^2) \label{a26}
\end{equation}
where  $x_{i}$ and $p_{j}$ satisfy the canonical commutation relations
$[x_{i},p_{j}]=i\hbar \delta_{ij}$.

Quantum gravitational properties have been connected with black hole physics by the generalized uncertainty principle. The influence of the quantum gravity to the black hole properties has been widely investigated by utilizing GUP. Some problems have been investigated in
\cite{BHT,BHE,KMTB,ADSG,RSRM,Chen:2013pra,Chen:2013tha}. In \cite{BHT}, the thermodynamics of a small black hole was discussed and shown that the Gross-Perry-Yaffe phase transition for a small black hole was interrupted by the minimal length induced by the GUP. The quantum corrected value of a Schwarzschild black hole entropy and a Reissner-Nordstrom black hole with double horizon were calculated by utilizing the proposed GUP\cite{BHE}. The thermodynamic quantities and the stability of a black hole in a cavity were studied and the result shown that the small black hole was unstable\cite{KMTB}. The derivation of the Hawking temperature of a Schwarzschild black hole was extended to the de Sitter and anti-de Sitter spacetimes\cite{ADSG}. The remnant mass and the corrected  mass-temperature relation, area law and heat capacity were obtained in \cite{RSRM}. In\cite{Chen:2013pra,Chen:2013tha}, a modified Dirac equation in curved spacetime and corrected Hawking temperatures had been derived. Though our works are similar to theirs, we will discuss that we has adopted different modified scheme and obtained different result clearly.

In this paper, we will investigate fermion's tunnelling process cross the 4-dimensional
Schwarzschild black hole's event horizon under the effect of the quantum gravity. we first derive the modified Dirac equation in Minkowski spacetime and subsequently generalize it into curved spacetime. The influence of the quantum gravity is introduced by GUP and the Dirac energy operator which modified by the minimal length scenario. We use the Hamilton-Jacobi method to solve the equation of motion of the spinor field. Then
the tunnelling rate and Hawking temperature are calculated. The results indicate that the corrected Hawking temperature depends on both the black hole's mass and the mass  of emitted fermions. Moreover, though many literatures imply that the existence of the minimal length and GUP will prevent the black hole's evaporation at some point, our results really show that the emerge of the quantum gravity will accelerate the black hole's evaporation.

The organization of the rest paper is as follows. In section 2, we generalize the dirac equation by adopting the quantum gravity effect. In section 3, the fermion's tunnelling behavior is investigated and the corrected Hawking temperature is obtained. The last section is the discussion and conclusion.
\section{Generalize the Dirac equation from Minkowski spacetime to curved spacetime}
\label{sec:1}
The Dirac equation in the ordinary quantum mechanics is
\begin{equation}
\label{edirace}
i\hbar \frac{\partial | \psi \rangle}{\partial t} = \left[ c \left(\vec{\alpha} \cdot \hat{\bf p} \right) + \hat{\beta}mc^{2} \right]| \psi \rangle ,
\end{equation}
where
\begin{equation}
\hat{\beta} =
\begin{pmatrix}
1 & 0 \cr
0 & -1 \cr
\end{pmatrix},
\end{equation}
\begin{equation}
\vec{\alpha} =
\begin{pmatrix}
0 & \vec{\sigma} \cr
\vec{\sigma} & 0 \cr
\end{pmatrix},
\end{equation}
and $\vec{\sigma}$ are the Pauli matrices. So it is easy to obtain the energy operator as
\begin{equation}
\hat{E} = c \left(\vec{\alpha} \cdot \hat{\bf p} \right) + \hat{\beta}mc^{2}.
\end{equation}

 To take the  quantum gravity into account, we replace $\hat{p_i}$ by $\hat{P_i}$ directly. Thus the new energy operator is
\begin{equation}
\label{rE1}
\hat{E}_{QG} = c \left(\vec{\alpha} \cdot \hat{\bf P} \right) + \hat{\beta}mc^{2}.
\end{equation}

Now, using the Eq. (\ref{a26}), $\hat{E}_{QG}$ can be rewriten as
\begin{equation}
\label{rE3}
\hat{E}_{QG}= c\left(\vec{\alpha} \cdot \hat{\bf p} \right) + \hat{\beta}mc^{2}  +	\beta c\left( \vec{\alpha} \cdot \hat{\bf p} \right)^{3},
\end{equation}
where we have used the relation $\hat{\bf p}^{2} = \left(\vec{\alpha} \cdot \hat{\bf p} \right)^{2}$.

 We should realize that just resembling the new momentum operator $\hat{\bf P}$ no longer coincides with the generator of space translation $-i\vec{\nabla}$ like (\ref{a26}), the new energy operator $\hat{E}_{QG}$ no longer coincides with the generator of time translation $i\frac{\partial}{\partial t}$ too \cite{ARU}. They are related by\cite{SMJS}
\begin{equation}
\label{Et}
\hat{E}_{QG} \equiv i \hbar \frac{\partial}{\partial t} \left( 1 + \beta \hbar^{2} \frac{\partial^{2}}{\partial t^{2}} \right).
\end{equation}
Then from (\ref{rE3}) and (\ref{Et}) we get
\begin{multline}
\label{Dirac1}
\left( i \hbar \frac{\partial}{\partial t} + i \beta \hbar^{3} \frac{\partial^{3}}{\partial t^{3}} \right) | \psi_{QG} \rangle \\=\left[ -i \hbar c\left(\vec{\alpha} \cdot \vec{\nabla} \right) + \hat{\beta}mc^{2}  +	i \beta \hbar^{3} c\left( \vec{\alpha} \cdot \vec{\nabla} \right)^{3} \right]
 | \psi_{QG} \rangle ,
\end{multline}
which treats space and time in a manifestly symmetric fashion. In Minkowski space-time, this is the generalized Dirac equation. And next we will again generalize it to curved space-time. \\
We set $G=c=1$.
The Latin indices are raised and lowered by flat metric $\eta_{ab}$ while Greek indices by curved metric $g_{\mu \nu}$. Then we can build the tetrad by
\begin{equation}
\begin{split}
& g_{\mu\nu}=e_\mu^{\ a}e_\nu^{\ b}\eta_{ab},\ \quad\eta_{ab}=g_{\mu\nu}e^\mu_{\ a}{e^\nu_{\ b}} \\ & e^\mu_{\ a}{e_\nu^{\ a}}=\delta^u_\nu \quad\quad , \ \ \  e^\mu_{\ a}{e_\mu^{\ b}}=\delta^a_b.
\end{split}
\end{equation}
So the equation (\ref{Dirac1}) can be rewritten as
\begin{equation}
\label{Dirac2}
\begin{split}
&\bigg[i\hbar\hat{\beta}\frac{\partial}{\partial{t}}+i\hbar(\hat{\beta}\vec{\alpha}\cdot\vec{\nabla})
+i\beta\hbar^3\hat{\beta}^3\frac{\partial^3}{\partial{t^3}}
-\\ &i\beta{\hbar^3}{(\hat{\beta}\vec{\alpha}\cdot\vec{\nabla}})^3-m\bigg] |\psi\rangle=0.
\end{split}
\end{equation}\\
We must note that the metric is
\begin{equation}
\eta^{ab}=diag\lbrace1,-1,-1,-1\rbrace.
\end{equation}
But if utilizing the convention in general relativity with  $\eta^{ab}=diag\lbrace-1,1,1,1\rbrace$,
we should make the transformation like
\begin{equation}
\begin{split}
ds^2&=dt^2-dx^2\\ &=-d(it)^2+d(ix)^2\\ &=-dt'^2+dx'^2, i.e.\;\;t'=it,x'^a=ix^b,\cdots
\end{split}
\end{equation}
\begin{equation}
\begin{split}
\{i\gamma^a,i\gamma^b\}&=-\{\gamma^a,\gamma^b\}\\ &=-2\eta^{ab}\\ &=2diag\{-1,1,1,1\}, i.e. \;\;
\gamma'^a=i\gamma^b
\end{split}
\end{equation}
It can be verified that the form of the equation (\ref{Dirac2}) is not changed after employing the transformation. After the transformation, we choose the gammas of flat spacetime as below
\begin{equation}
\label{b1}
\begin{split}
&\gamma^0=\begin{pmatrix}iI & 0\\0 & -iI\end{pmatrix}, \quad \ \ \
\gamma^1=\begin{pmatrix}0&i\sigma^3\\-i\sigma^3&0\end{pmatrix}, \quad \\
&\gamma^2=\begin{pmatrix}0&i\sigma^1\\-i\sigma^1&0\end{pmatrix}, \quad
\gamma^3=\begin{pmatrix}0&i\sigma^2\\-i\sigma^2&0\end{pmatrix}. \quad
\end{split}
\end{equation}
Then the  Lorentz spinor generators are defined by
\begin{equation}
\Sigma_{ab}=\frac{i}{4}\{\gamma^a,\gamma^b\},\quad\{\gamma^a,\gamma^b\}=2\eta^{ab}.
\end{equation}
And the ${\gamma^{\mu}}^{,}s$ in curved spacetime can be constructed as
\begin{equation}
\label{b2}
\gamma^\mu=e^\mu_{\ a}\gamma^a,\quad\{\gamma^\mu,\gamma^\nu\}=2g^{\mu\nu}.
\end{equation}
To get the generalized Dirac equation in curved spacetime, we rewrite eq. (\ref{Dirac2}) in a covariant form by replacing $\partial_\mu$ with $\partial_\mu+\Omega_\mu$ and $\gamma^{a}$'s with $\gamma^{\mu}$'s
\begin{equation}
\begin{split}
[i\hbar\gamma^\mu(\partial_\mu+\Omega_\mu)+&i\beta\hbar^3(\gamma^0\partial_0+\gamma^0\Omega_0)^3 i\beta\hbar^3\\ & -(\gamma^i\partial_i+\gamma^i\Omega_i)^3-m]|\psi\rangle=0\label{gde}
\end{split}
\end{equation}
where $\Omega_\mu\equiv\frac{i}{2}\omega_\mu^{\ ab}\Sigma_{ab}$ and the $\omega_{\mu}^{\  ab} $ is spin connection which can be written by the tetrad and ordinary connection as $\omega_{\mu\ b}^{\ a}=e_\nu^{\ a}\Gamma_{\mu\lambda}^\nu e^\lambda_{\ b}-e^\lambda_{\ b}\partial_\mu
e_\lambda^{\ a}$.
In the next section, we will solve this equation in Schwarzschild spacetime.

\section{Fermion's tunnelling in the Schwarzschild spacetime with quantum gravity correction }
\label{sec:2}
In this section, with the effects of quantum gravity taken into account, we discuss the spin-1/2 fermion's tunnelling behavior under the Schwarzschild black hole. The metric of Schwarzschild spacetime used in this paper is:
\begin{equation}
ds^2=-f(r)dt^2+\frac{1}{g(r)}dr^2+r^2(d\theta^2+sin^2\theta d\phi^2)
\end{equation}
with $f(r)=g(r)=1-\frac{2M}{r}$, and M is the black hole's mass. The event horizon is located at $r_h=2M$. The tetrad of the metric is constructed as
\begin{align}
e_\mu^{\ a}&=diag(\sqrt{f(r)},\frac{1}{\sqrt{g(r)}},\frac{1}{\sqrt{g^{\theta\theta}}},\frac{1}{\sqrt{g^{\phi\phi}}})\nonumber\\
&=diag(\sqrt{f},1/\sqrt{g},r,rsin\theta)
\end{align}
From (\ref{b1}) and (\ref{b2}), the gamma matrices in Schwarzschild space-time are given as
\begin{align}
\begin{split}
&\gamma^t=e^t_{\  0}\gamma^0=\frac{1}{\sqrt{f(r)}}\begin{pmatrix}iI&0\\0&-iI\end{pmatrix}\\
&\gamma^r=e^r_{\  1}\gamma^1=\sqrt{g(r)}\begin{pmatrix}0&i\sigma^3\\ -i\sigma^3&0\end{pmatrix}\\
&\gamma^\theta=e^\theta_{\  2}\gamma^2=\sqrt{g^{\theta\theta}}\begin{pmatrix}0&i\sigma^1\\ -i\sigma^1&0\end{pmatrix}\\
&\gamma^\phi=e^\phi_{\  3}\gamma^3=\sqrt{g^{\phi\phi}}\begin{pmatrix}0&i\sigma^2\\
-i\sigma^2&0\end{pmatrix}
\end{split}.\label{a16}
\end{align}
The fermions motion is governed by equation (\ref{gde}). For a spin-1/2 particle, the wave function of the spin up state is assumed semi-classiclly as
\begin{equation}
|\psi\rangle=\begin{pmatrix}
A\\0\\B\\0\end{pmatrix}exp\left(\frac{i}{\hbar}I(t,r,\theta,\phi)\right) \label{a9}
\end{equation}
where A, B and I are functions of coordinates $t, r, \theta, \phi$, and I is the action of the emitted fermions. The discussing of the spin down state is the same as that of the spin up. We just consider the spin up state in this paper. \\
To solve the generlized Dirac equation (\ref{gde}) with employing the WKB approximation, we substitute (\ref{a16})(\ref{a9}) into (\ref{gde}) and neglect the terms including $\partial A$, $\partial B$ and high orders of $\hbar$. If solving (\ref{gde}) by neglecting the high order items of $\hbar$, the $\Omega_\mu $ can be neglect and only the term with the partial operator's order is just 3 can be survived. \\
Then we get the simplified equation respect to (\ref{gde})
\begin{equation}
\left[\hbar\gamma^\mu\partial_u+\beta\hbar^3(\gamma^0\partial_0)^3-\beta\hbar^3(\gamma^i\partial_i)^3+im\right]|\psi\rangle=0\label{a6}
\end{equation}
and finally obtain decoupled four Hamilton-Jacobi equations
\begin{equation}
\begin{split}
A[-\frac{1}{\sqrt{f}}\partial_t I-&\beta\frac{1}{f\sqrt{f}}(\partial_t I)^3+im]- B \{1+\beta[g^{rr}(\partial_r I)^2+\\ &g^{\theta\theta}(\partial_{\theta}I)^2+g^{\phi\phi}(\partial_{\phi}I)^2] \}\sqrt{g^{rr}}\partial_r I=0 \label{a12}
\end{split}
\end{equation}
\begin{equation}
\begin{split}
-B\{1+& \beta[g^{rr}(\partial_r I)^2+g^{\theta\theta}(\partial_{\theta}I)^2+g^{\phi\phi}(\partial_{\phi}I)^2] \}\\ &(\sqrt{g^{\theta\theta}}\partial_{\theta} I+i\sqrt{g^{\phi\phi}}\partial_{\phi} I)=0\label{a13}
\end{split}
\end{equation}
\begin{equation}
\begin{split}
A\{1+&\beta[g^{rr}(\partial_r I)^2+g^{\theta\theta}(\partial_{\theta}I)^2+g^{\phi\phi}(\partial_{\phi}I)^2] \}\sqrt{g^{rr}}\partial_r I \\ &+B[\frac{1}{\sqrt{f}}\partial_t I +\beta\frac{1}{f\sqrt{f}}(\partial_t I)^3+im]=0\label{a14}
\end{split}
\end{equation}
\begin{equation}
\begin{split}
A\{1+&\beta[g^{rr}(\partial_r I)^2+g^{\theta\theta}(\partial_{\theta}I)^2+g^{\phi\phi}(\partial_{\phi}I)^2] \}\\ &(\sqrt{g^{\theta\theta}}\partial_{\theta} I+i\sqrt{g^{\phi\phi}}\partial_{\phi} I)=0\label{a15}
\end{split}
\end{equation}
To solve the Hamilton-Jacobi equations above, the action is set as below to separate the variables
\begin{equation}
I=-\omega t+W(r)+\Theta(\theta,\phi)\label{a11}
\end{equation}
where $\omega$ is the energy of the emitted particle.
Substitute (\ref{a11}) into [\ref{a12}-\ref{a15}], we gain four equations below
\begin{equation}
\begin{split}&A[\frac{\omega}{\sqrt{f}}+\beta\frac{\omega^3}{f\sqrt{f}}+im]- B \{1+\\ &\beta[g^{rr}(\partial_r W)^2+g^{\theta\theta}(\partial_{\theta}\Theta)^2+g^{\phi\phi}(\partial_{\phi}\Theta)^2] \}\sqrt{g^{rr}}\partial_r W=0\label{a17}\end{split}
\end{equation}
\begin{equation}
\begin{split}
-B\{1+&\beta[g^{rr}(\partial_r W)^2+g^{\theta\theta}(\partial_{\theta}\Theta)^2+g^{\phi\phi}(\partial_{\phi}\Theta)^2] \}\\&(\sqrt{g^{\theta\theta}}\partial_{\theta}\Theta+i\sqrt{g^{\phi\phi}}\partial_{\phi} \Theta)=0\label{a18}\end{split}
\end{equation}
\begin{equation}
\begin{split}&A\{1+\beta[g^{rr}(\partial_r W)^2+g^{\theta\theta}(\partial_{\theta}\Theta)^2+g^{\phi\phi}(\partial_{\phi}\Theta)^2] \}\sqrt{g^{rr}}\partial_{r} W \\ &+B[-\frac{\omega}{\sqrt{f}} -\beta\frac{\omega^3}{f\sqrt{f}}+im]=0\label{a19}\end{split}
\end{equation}
\begin{equation}
\begin{split}A\{1+&\beta[g^{rr}(\partial_r W)^2+g^{\theta\theta}(\partial_{\theta}\Theta)^2+g^{\phi\phi}(\partial_{\phi}\Theta)^2] \}\\ &(\sqrt{g^{\theta\theta}}\partial_{\theta} \Theta+i\sqrt{g^{\phi\phi}}\partial_{\phi} \Theta)=0\label{a20}\end{split}
\end{equation}
Contrasting (\ref{a18}) and (\ref{a20}), we find they are identical except the letters A and B. We rewrite them as
\begin{equation}
\begin{split}
\{1+&\beta[g^{rr}(\partial_r W)^2+g^{\theta\theta}(\partial_{\theta}\Theta)^2+g^{\phi\phi}(\partial_{\phi}\Theta)^2] \}\\ &(\sqrt{g^{\theta\theta}}\partial_{\theta} \Theta+i\sqrt{g^{\phi\phi}}\partial_{\phi} \Theta)=0.
\end{split}
\end{equation}
For $\beta$ is a small quantity repesenting the effects of quantum gravity, the part in the brace bracket can not be zero. And we know it from previous work that $\Theta$ has no contribution to the tunnelling rate. Then there must be
\begin{equation}
\sqrt{g^{\theta\theta}}\partial_{\theta} \Theta+i\sqrt{g^{\phi\phi}}\partial_{\phi} \Theta=0.
\end{equation}
The modulus of it is
\begin{equation}
(\sqrt{g^{\theta\theta}}\partial_{\theta} \Theta)^2+(\sqrt{g^{\phi\phi}}\partial_{\phi} \Theta)^2=0. \label{a21}
\end{equation}
Then substituting (\ref{a21}) into (\ref{a17}) and (\ref{a19}), a group of linear equation with respecting to A and B is attained
\begin{multline}
A[\frac{\omega}{\sqrt{f}}+\beta\frac{\omega^3}{f\sqrt{f}}+im]-\\B [1+\beta g^{rr}(\partial_r W)^2]\sqrt{g^{rr}}\partial_{r} W=0, \nonumber
\end{multline}
\begin{multline}
A[1+\beta g^{rr}(\partial_r W)^2]\sqrt{g^{rr}}\partial_r W+\\B[-\frac{\omega}{\sqrt{f}} -\beta\frac{\omega^3}{f\sqrt{f}}+im]=0
\end{multline}
If expecting none zero solutions, its determinant must be zero.
Then a differential equation of W is gained like
\begin{equation}
-(\frac{\omega}{\sqrt{f}}+\beta\frac{\omega^3}{f\sqrt{f}})^2-m^2+[1+\beta g^{rr}(\partial_{r} W)^2]^2 g^{rr}(\partial_r  W)^2=0.\label{a24}
\end{equation}
Neglecting the high order of $\beta$ and expanding (\ref{a24}), W is assumed as
\begin{equation}
W(r)=W_0+\beta W_1\label{a23}.
\end{equation}
Inserting (\ref{a23}) into (\ref{a24}) and considering $g=f=1-\frac{2M}{r}$, we get two differential equations
\begin{align}
&\pm\sqrt{m^2+\frac{\omega^2}{g}}+\sqrt{g}\partial_r W_0=0  \label{b3} \\
&\pm\frac{\omega^4}{g^2\sqrt{m^2+\frac{\omega^2}{g}}}+g^{\frac{3}{2}}({\partial_r W_0})^3+\sqrt{g}\partial_r W_1=0
\end{align}
Solving the above two equations at the event horizon, the result is
\begin{equation}
W=\pm i2\pi M\omega(1-2m^2\beta).
\end{equation}
The $+/-$ sign corresponds to outgoing/ingoing wave.
If just from (\ref{b3}), we recover the original outcome without any correction
\begin{equation}
W_0=\pm i2\pi M\omega.
\end{equation}
Then the tunnelling rate\cite{PMTF} of the spin-$\frac{1}{2}$ fermion crossing the horizon is
\begin{equation}
\begin{split}
\Gamma &= \frac{P_(emission))}{P_(absorption)}\\&=\frac{exp(-2ImI_+)}{exp(-2ImI_-)}\\&=\frac{-2ImW_+ -2Im\Theta}{-2ImW_- -2Im\Theta}\\
&=exp[-8\pi M\omega(1-2m^2\beta)].\\
\end{split}
\end{equation}
Finally, the corrected Hawking temperature is
\begin{align}
T&=\frac{1}{8\pi M(1-2m^2\beta)} \nonumber\\
&=(1+2m^2\beta)T_0 \label{b4}
\end{align}
where $T_0=\frac{1}{8\pi M} $ is the original Hawking temperature.

For massless particle tunnelling in the schwarzschild black hole, the corrected Hawking temperature was derived in \cite{MPKS}. Respecting to our job, a similar work has been done in \cite{Chen:2013pra,Chen:2013tha}. The corrected Hawking temperature corresponding to different black hole were calculated in these paper. But their Hawking temperature respecting to the schwarzschild black hole
\begin{align}
T&=\frac{1}{8\pi M\left(1+\frac{1}{2}\beta(3m^2+4\omega^2)\right)} \nonumber\\
&=\left[1-\frac{1}{2}\beta(3m^2+4\omega^2)\right]T_0 \label{b5}
\end{align}
is different from ours, where $T_0=\frac{1}{8\pi M} $ is the original Hawking temperature.
Comparing (\ref{b4}) with (\ref{b5}), the correction value of Hawking temperature in our outcome do not rely on the radiated fermion particle's energy $\omega$. And the even more important distinction is the sign in the result. The minus sign means the effects of quantum gravity will retard the black hole evaporation and the positive means the effects of quantum gravity will accelerate the black hole evaporation. So according to our result, there is no remnant at last.\\
Actually, the utilizing of different energy operator which according to GUP leads to different modified scheme and result. The below energy operator was adopted in \cite{Chen:2013pra}
\begin{equation}
i\hbar\partial_t\equiv\hat{E}_{QG}\left(1-\frac{\beta}{c^2}\hat{E}^2_{QG}\right)
\end{equation}
and we adopt another one (\ref{Et}). So we have used a distinct modified scheme and obtained a different result.

\section{Discussion and conclusion }
\label{sec:3}
In this paper, we first modified the Dirac equation in Minkowski spacetime with employing the generalized energy and momentum operator and original Dirac equation. Then, we made a transformation to generalize it into curved spacetime. By assuming the existence of minimal length, the effects of quantum gravity was introduced via generalized uncertainty principle through the modification process. Finally, the radiation of spin 1/2 particles in the 4-dimensional Schwarzschild spacetime was calculated with Hamilton-Jacob method. We got the corrected tunnelling rate and Hawking temperature.

We found that the quantum corrected part of Hawking temperature is depend on both the black hole mass M and the radiated particle mass m. Though many literatures imply the quantum gravity correction will retard the black hole evaporation and lead to the existence of remnants, our result really indicates that the influence of quantum gravity will accelerating black hole evaporation and the remnants can not be exist.

We just offered a possible modification scheme to investigate fermion's tunnelling from schwarzschild black hole. We only calculated the leading order of $\hbar$ and $\beta$. As for more complex black hole and high orders of corrections, it is expected to be studied in the future work.

\section{Acknowledgements}
\label{sec:4}
 Authors are very grateful for Ran Li and ShaoWen Wei's discussion. This work was supported by the Fundamental Research Funds for the Central Universities No.lzujbky-2013-16.

%

\end{document}